# Origin of superconductivity and latent charge density wave in $NbS_2$

Christoph Heil,[1] Samuel Poncé,[1] Henry Lambert,[1,*] Martin Schlipf,[1] Elena R. Margine,[2] and Feliciano Giustino[1,†]

[1]Department of Materials, University of Oxford, Parks Road, Oxford OX1 3PH, United Kingdom

[2]Department of Physics, Applied Physics and Astronomy, Binghamton University-SUNY, Binghamton, New York 13902, USA

We elucidate the origin of the phonon-mediated superconductivity in $2H$-$NbS_2$ using the *ab initio* anisotropic Migdal-Eliashberg theory including Coulomb interactions. We demonstrate that superconductivity is associated with Fermi surface hot spots exhibiting an unusually strong electron-phonon interaction. The electron-lattice coupling is dominated by low-energy anharmonic phonons, which place the system on the verge of a charge density wave instability. We also provide definitive evidence for two-gap superconductivity in $2H$-$NbS_2$, and show that the low- and high-energy peaks observed in tunneling spectra correspond to the $\Gamma$- and $K$-centered Fermi surface pockets, respectively. The present findings call for further efforts to determine whether our proposed mechanism underpins superconductivity in the whole family of metallic transition metal dichalcogenides.

Transition metal dichalcogenides (TMDs) have generated considerable interest in recent years, since they provide an ideal playground for studying semiconductors, metals, and superconductors in two dimensions using the same structural template [1–3]. In the case of superconducting TMDs, one remarkable feature is that Cooper pair condensation usually coexists with a charge density wave (CDW) [4], raising the question on whether superconductivity and CDW co-operate or compete in these compounds [5–12].

Within the family of superconducting TMDs, $2H$-$NbS_2$ stands out as the only system for which a CDW phase has not been observed [13, 14]. This suggests that a comparative analysis of $NbS_2$ and other superconducting TMDs may help to clarify the interplay between the superconductive and the CDW instabilities in the entire family. $2H$-$NbS_2$ is a phonon-mediated superconductor with a critical temperature $T_c = 5.7$ K. Scanning tunneling spectroscopy (STS) measurements on this compound revealed two pronounced features in the density of states (DOS) at 0.53 meV and 0.97 meV below the critical temperature, providing strong indications of two-gap superconductivity [14]. However, so far microscopic calculations have considered only a single-gap scenario [15, 16].

In this work we investigate the nature of the superconducting gap and the pairing mechanism in $2H$-$NbS_2$ using the fully anisotropic *ab initio* Migdal-Eliashberg theory, and describe both electron-phonon and electron-electron interactions without any adjustable parameters. Our key finding is that a very significant contribution to the superconducting pairing comes from the low-energy anharmonic phonons with wavevectors near the line connecting the $M$ and $L$ points. These are the same phonons responsible for the CDW instability in other TMDs [8, 11, 17–19], indicating that superconductivity in $NbS_2$ is intimately connected with a latent CDW. In agreement with the STS experiments of Ref. 14, we find two distinct and anisotropic superconducting gaps.

All calculations reported in this work were performed using density functional theory (DFT) in the local density approximation [20, 21]. We employed the Quantum Espresso package [22] for the electronic structure and lattice dynamics, the Epw code [23–25] for the electron-phonon interaction (EPI) and the superconducting gap, the SternheimerGW code [26, 27] for the Coulomb interaction and for GW quasiparticle calculations of the band structure, and the Wannier90 code [28] for generating maximally-localized Wannier functions [29].

$2H$-$NbS_2$ crystallizes in a layered hexagonal structure (space group $P6_3/mmc$), with the unit cell containing two S-Nb-S sandwiches [30]. Figure 1 shows that the Fermi surface of $2H$-$NbS_2$ consists of three distinct sheets: (i) a disc-shaped hole pocket centered at $\Gamma$ ($S_{\Gamma_1}$), of predominant S-$p_z$ character; (ii) a cylindrical hole pocket also centered at $\Gamma$ ($S_{\Gamma_2}$), arising from Nb-$d_{z^2}$ orbitals; (iii) a triangular hole pocket centered around $K$ ($S_K$) which originates from Nb-$d_{z^2}$ states near the $M$ point, and from $d_{xy}$ and $d_{x^2-y^2}$ states near $K$.

In order to calculate the superconducting gap and critical temperature of $2H$-$NbS_2$, we start from the screened Coulomb interaction. Coulomb effects are included in the Migdal-Eliashberg equations via the Morel-Anderson pseudopotential $\mu^* = \mu/[1 + \mu \log(\omega_{el}/\omega_{ph})]$ [31]. In this expression $\omega_{el}$ and $\omega_{ph}$ are characteristic electron and phonon energies, respectively [32], and $\mu$ is a double Fermi-surface average of the screened Coulomb interaction $W$: $\mu = N_F\langle\langle\, \langle \mathbf{k}, -\mathbf{k}|W|\mathbf{k}', -\mathbf{k}'\rangle\, \rangle\rangle_{FS}$, where $N_F$ is the DOS at the Fermi level and $W$ is calculated within the random phase approximation [26, 27, 33, 34]. Our calculations yield $\mu^* = 0.20$ ($\mu = 0.37$), which is larger than typical values found in isotropic superconductors. This large Morel-Anderson pseudopotential originates from very strong repulsive interactions at a few Fermi surface hot spots. The electron-electron repulsion is weakest for the states corresponding to in-plane orbitals, that is in the center of the Fermi surface sheet $S_K$

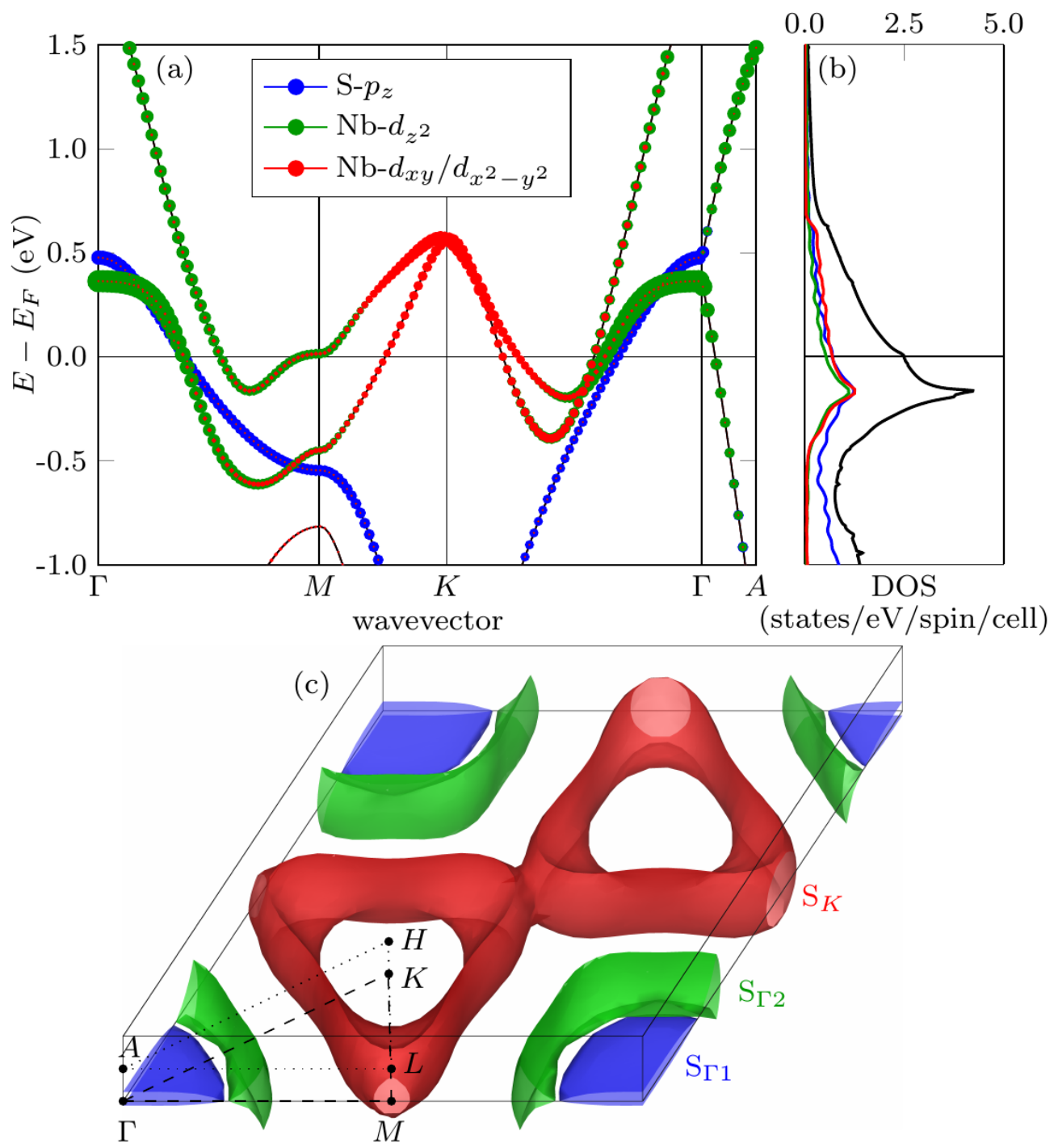

Figure 1. (Color online) (a) Calculated band structure of $2H$-$NbS_2$, with the orbital contributions proportional to the size of the colored dots, as indicated in the legend: S-$p_z$ (blue), Nb-$d_{z^2}$ (green) and Nb-$d_{xy}/d_{x^2-y^2}$ (red). (b) Corresponding DOS and decomposition into atomic orbitals, using the same color code as in (a). The black line is the total DOS. (c) Fermi surface showing the three distinct Fermi surface sheets labeled as $S_{\Gamma_1}$ (blue), $S_{\Gamma_2}$ (green), and $S_K$ (red); as well as the high-symmetry points in the Brillouin zone.

[see Fig. 1(c) and Supplemental Fig. S1 [35]].

Figures 2(a) and (b) show the calculated phonon dispersion relations of $2H$-$NbS_2$. In the harmonic approximation the two lowest-energy vibrational modes near the $M$ and $L$ points of the Brillouin zone have imaginary frequencies, as shown in Supplemental Fig. S2(f)-(g) [35]. For each of these modes we performed fully anharmonic calculations by mapping the DFT potential energy surface, and used the renormalized anharmonic frequencies in all calculations of phonon dispersions and EPIs. The resulting dispersion relations are in good agreement with inelastic X-ray scattering experiments [19]. At the $M$ point the anharmonic modes have $A_g$ and $B_u$ symmetry, respectively. The DFT potential energy surfaces of these modes correspond to symmetric double-wells, and can be identified with the modes which drive the CDW instability in related TMDs [8, 11, 17, 18]. Details about these calculations and comparisons with other methods [36–42] are provided in the Supplemental Material [35].

In Fig. 2(d)-(e) we show the wavevector- and mode-resolved electron-phonon coupling strength $\lambda^{\rm ph}_{\mathbf{q},\nu}$ along high-symmetry lines. We see that the two anharmonic branches exhibit anomalously large EPIs (up to $\lambda^{\rm ph}_{\mathbf{q},\nu=1} = 23.5$). In order to check the weight of these modes on the total EPI, we show in Fig. 2(c) the isotropic Eliashberg function $\alpha^2F(\omega)$ [25] separated into contributions arising from the two low-energy anharmonic modes (green and red); and the remaining modes (blue). The corresponding breakdown of the total EPI strength [$\lambda = 2\int_0^\infty \alpha^2F(\omega)/\omega \; d\omega$] is shown by the dashed lines; this analysis indicates that the anharmonic modes contribute more than 50% to the total interaction strength. At variance with these modes, the contributions of all other modes are relatively homogeneous and follow the vibrational DOS [Fig. 2(c)].

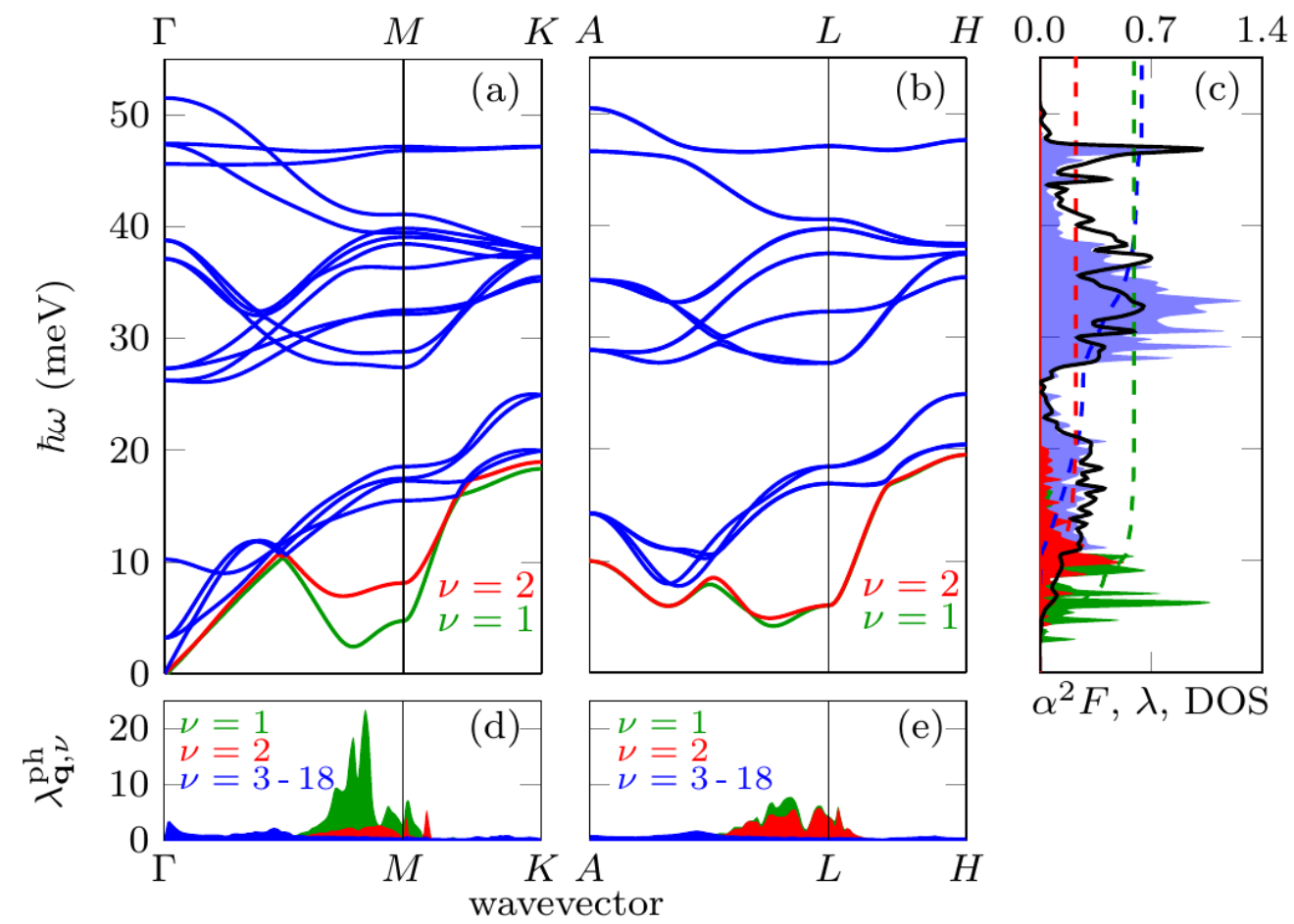

Figure 2. (Color online) (a)-(b) Calculated phonon dispersion relations of $2H$-$NbS_2$ for the first mode ($\nu = 1$, green), the second mode ($\nu = 2$, red), and modes $\nu = 3 - 18$ (blue). (c) Calculated vibrational DOS (solid, black line), isotropic Eliashberg function $\alpha^2F(\omega)$ (shaded areas) and cumulative EPI strength $\lambda(\omega)$ (dashed lines). The blue curves are for modes $3 - 18$, and the green and red curves are for the two lowest-energy phonons. (d)-(e) The mode-resolved EPI strength $\lambda^{\rm ph}_{\mathbf{q},\nu}$ for the phonon modes in (a)-(b).

Anomalous EPI strengths of selected phonons can arise either from Fermi surface nesting effects [43], or from the breaking of electronic degeneracies by lattice fluctuations and the consequent removal of electronic weight from the DOS close to the Fermi level [44], in analogy to the dynamical Jahn-Teller effect in molecules [45]. To determine which mechanism is active in the case of $2H$-$NbS_2$, we calculated the Fermi surface nesting function [25, 46]. By inspecting this function in Supplemental Fig. S3 [35] we see that there are no obvious nesting vectors in this system, likely because the sides of the triangular Fermi pocket $S_K$ are bulging inwards (Fig. 1). Therefore we rule out nesting as a possible cause of strong EPIs in the anharmonic modes. This is in line with previous reports on other TMDs [7–9, 47]. In order to test the second mechanism, we focused on the $B_u$ mode at the $M$ point. We doubled the $NbS_2$ unit cell along the Γ$M$ direction so as to fold $M$ into Γ, and calculated band structures, DOS and Fermi surfaces with or without a frozen $B_u$ phonon [48]. This phonon induces avoided crossings near the Fermi level in regions which correspond to the centers of the triangular Fermi arcs $S_K$ in Fig. 1(c). The deformation of the band structure is accompanied by a suppression of large parts of the Fermi surface in the supercell [as shown in Fig. 3 and Supplemental Fig. S4 [35]] and leads to the removal of the pronounced shoulder in the DOS close to the Fermi energy of the undistorted structure [see Fig. 1 and Supplemental Fig. S5 [35]]. As pointed out in Refs. 44 and 49, such changes in the electronic structure are indicative of a phonon softening and a latent lattice instability, and is in line with our findings of Fermi-surface EPI hot spots precisely in the middle of the triangular sides of $S_K$.

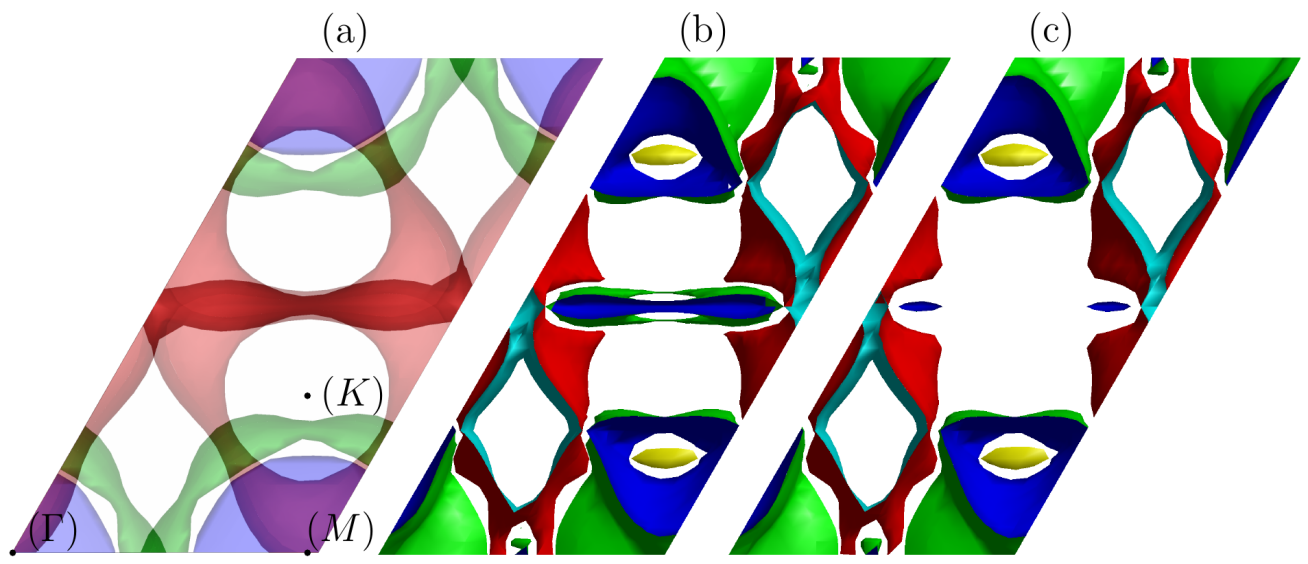

Figure 3. (Color online) (a) Sketch of the folded FS of the 2×1×1 supercell, obtained by overlaying two FSs of the crystal unit cell. The high-symmetry points are indicated in brackets since they refer to the crystal Brillouin zone. (b) Calculated FS of the 2×1×1 supercell in the ground-state structure. (c) Calculated FS of the 2×1×1 supercell after displacing the atoms according to the $B_u$ phonon mode at $M$, so as to place the structure in one of the minima of the double-well potential.

We now move to the superconducting properties of 2$H$-$NbS_2$. Figure 4(a) shows the distribution of temperature-dependent superconducting gaps on the Fermi surface, as calculated using the anisotropic *ab initio* Migdal-Eliashberg theory. Our calculations using the DFT band structure yield a critical temperature $T_c = 18.6$ K and a maximum zero-temperature superconducting gap $\Delta = 4.2$ meV, overestimating the experimental values of 5.7 K and 0.97 meV, respectively [14]. The origin of this discrepancy will be discussed further down; for now, in order to facilitate comparison with experiment, we apply the empirical scaling factors 5.7/18.6 and 0.97/4.2, respectively.

The superconducting gaps in Fig. 4(a) are seen to follow the standard BCS-type temperature dependence. At each temperature the gaps cluster around two distinct values, indicating that 2$H$-$NbS_2$ is a two-gap superconductor. The smaller gap is associated with the Fermi sheets $S_{\Gamma_1}$ and $S_{\Gamma_2}$, while the larger gap belongs to $S_K$. From the gaps we calculate the superconducting DOS, and in Fig. 4(b) we compare our results to the tunneling experiments of Ref. 14. The agreement between our calculations and experiments is very good (apart from the empirical scaling discussed above), and confirms that the peak around 1.0 meV and the shoulder around 0.6 meV in the tunneling data taken at 1.8 K are to be associated with two distinct superconducting gaps on the Γ-centered and on the $K$-centered Fermi surface pockets. Our finding of two distinct superconducting gaps is also in line with the anomalous temperature dependence of the specific heat [50] and the pressure dependence of the upper critical field [51].

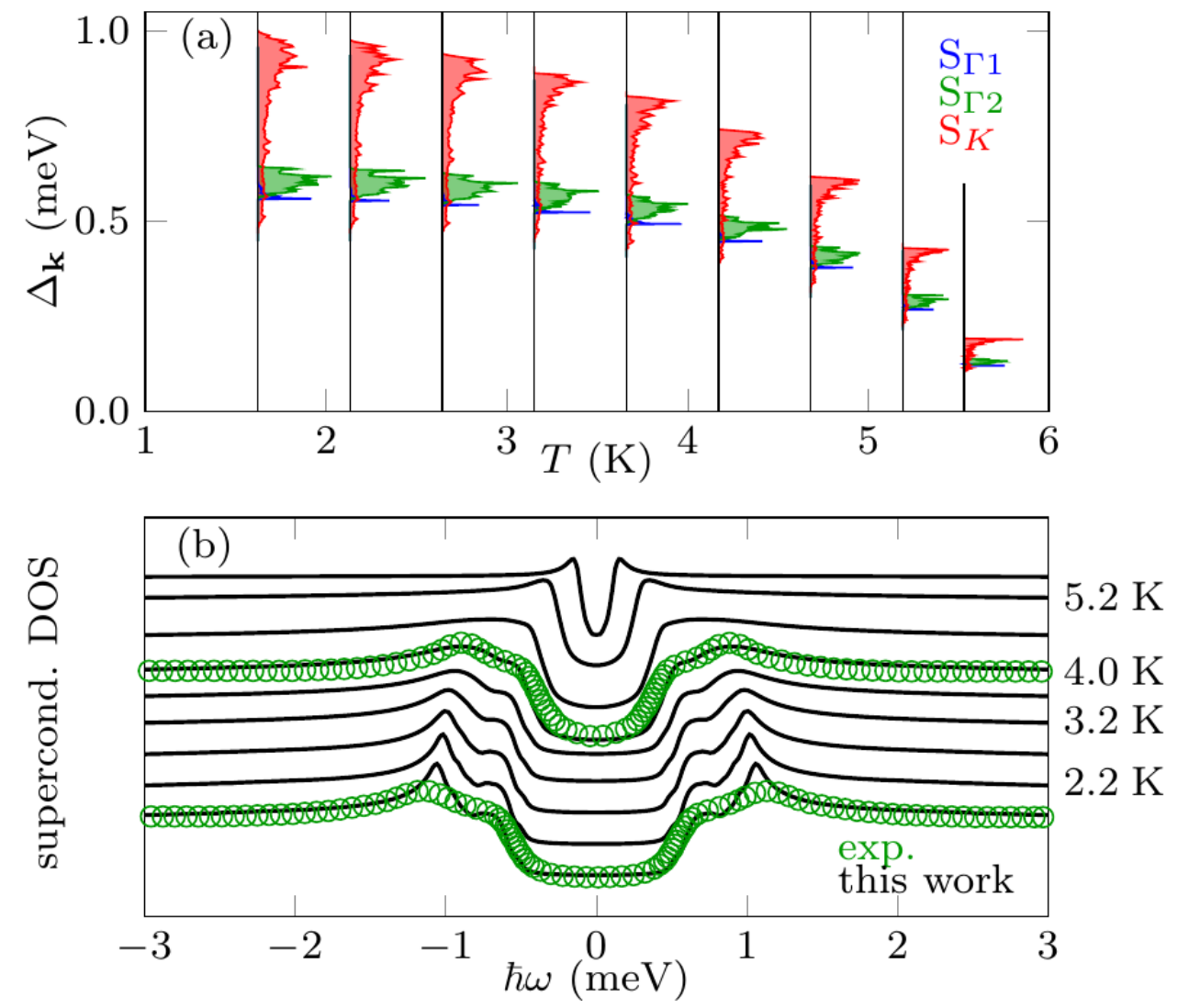

Figure 4. (Color online) (a) Energy distribution of the superconducting gap $\Delta_{\mathbf{k}}$ of 2$H$-$NbS_2$ as a function of temperature, calculated within the anisotropic *ab initio* Migdal-Eliashberg theory including Coulomb interactions. The colors indicate data belonging to different Fermi surface sheets. (b) Calculated DOS in the superconducting phase of 2$H$-$NbS_2$ (black lines), compared with tunneling data from Ref. 14 (green circles), for $T = 1.8$ K and $T = 4.0$ K. In both panels, to facilitate comparison with experiment, *the theoretical temperature and gaps were scaled by the factors* 5.7/18.6 *and* 0.97/4.2, *respectively.*

Figure 5(a) shows the momentum-resolved superconducting gap on the Fermi surface. We observe finite values for $\Delta_{\mathbf{k}}$ on the whole FS, therefore the FS is fully gapped below the critical temperature. The gaps on the $S_{\Gamma_1}$ and $S_{\Gamma_2}$ sheets exhibit narrow distributions centered around 0.57 meV and 0.56-0.65 meV, respectively. In contrast, the gap on the $S_K$ sheet is highly anisotropic and varies over the wide range 0.5-1.0 meV [Fig. 5(b)]. By recalling the orbital analysis of the Fermi surface in Fig. 1 we conclude that low values of the superconducting gaps are found on those regions of the Fermi surface with out-of-plane orbital character (S-$p_z$ and Nb-$d_{z^2}$), while large values correspond to regions with in-plane character (Nb-$d_{xy}$ and $d_{x^2-y^2}$).

The regions of the Fermi surface with the largest su-perconducting gap coincide with the hot spots of the electronic EPI parameter $\lambda^{\rm el}_{\mathbf{k}}$, as shown in Fig. 5(c). In particular the EPI strength on the Γ-centered pockets exhibits a narrow distribution around $\lambda^{\rm el}_{\mathbf{k}} = 0.8$-$0.9$, while that on the $K$-centered pocket covers the wide range from $\lambda^{\rm el}_{\mathbf{k}} = 0.6$ to $\lambda^{\rm el}_{\mathbf{k}} = 3.0$. The resulting average EPI parameter is $\lambda^{\rm el} = 1.46$. We emphasize that, while $\lambda^{\rm ph}_{\mathbf{k},\nu}$ (Fig. 2) and $\lambda^{\rm el}_{\mathbf{k}}$ (Fig. 5) are related, they are not the same quantity. A detailed discussion of these quantities can be found in Ref. 52. From this analysis it is clear that the two anharmonic modes contribute significantly to creating electron-phonon hot spots on the triangular arcs of the Fermi surface. While no charge ordering has been observed in $2H$-$NbS_2$ thus far, our findings clearly indicate that the EPI with the anharmonic phonons places this system very close to a lattice instability [53]. This conclusion is further supported by the Fermi surface contribution to the adiabatic phonon self-energy [54] presented in Supplemental Fig. S6 [35]. Based on these considerations we propose that superconductivity in $2H$-$NbS_2$ is driven by the same EPI that underlies a 'latent' CDW instability, i.e. a CDW that is possibly quenched by quantum fluctuations.

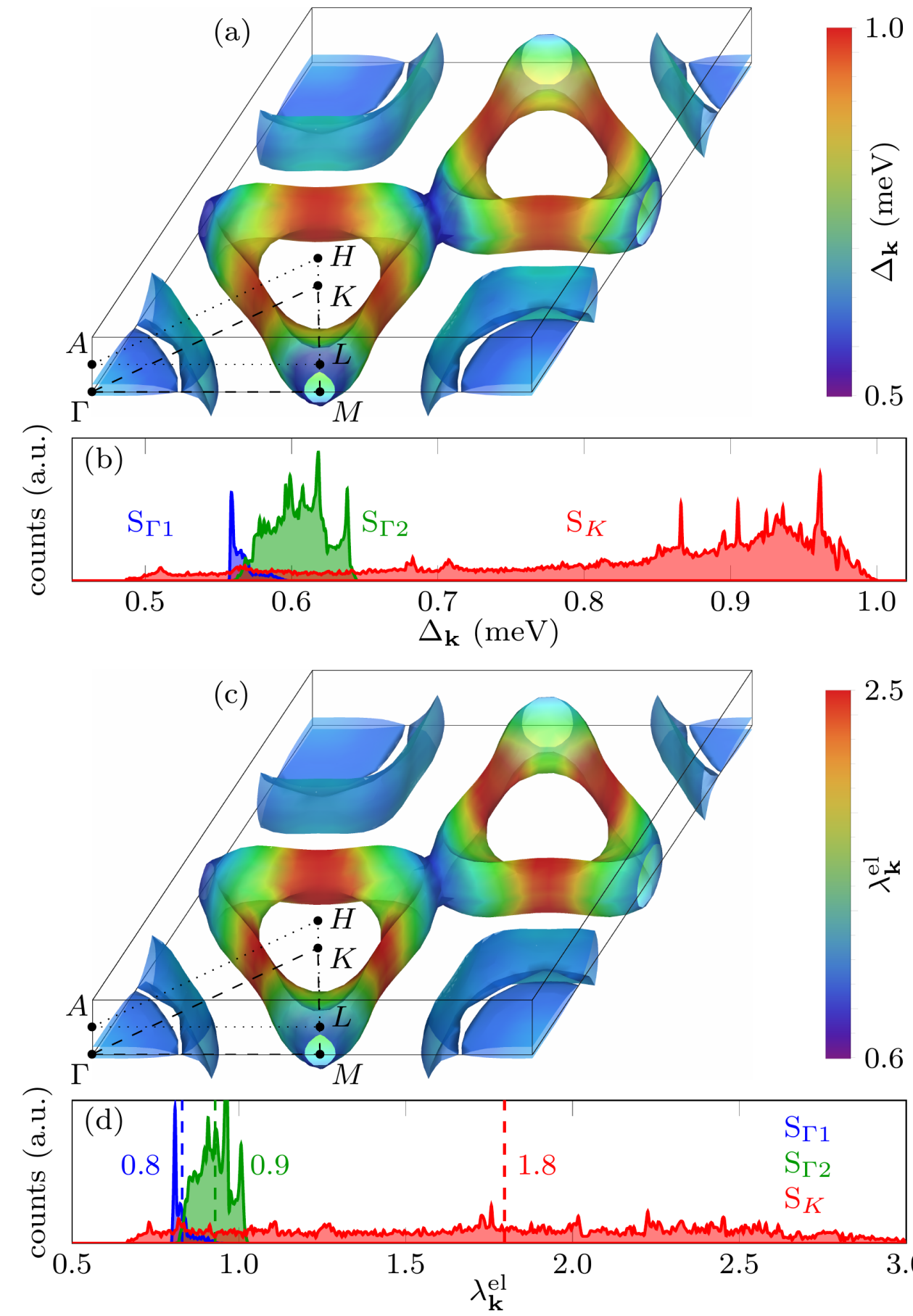

Figure 5. (Color online) (a) Momentum-resolved superconducting gap $\Delta_{\mathbf{k}}$ on the Fermi surface of $2H$-$NbS_2$, calculated within the *ab initio* Migdal-Eliashberg theory for $T = 1.7$ K. (b) Energy distribution of the superconducting gap $\Delta_{\mathbf{k}}$, color-coded by Fermi surface sheet: $S_{\Gamma_1}$ (blue), $S_{\Gamma_2}$ (green), and $S_K$ (red). The values for the gaps were scaled as in Fig. 4. (c) Momentum-resolved EPI strength $\lambda^{\rm el}_{\mathbf{k}}$ on the Fermi surface of $2H$-$NbS_2$. (d) Distribution of the EPI strength $\lambda^{\rm el}_{\mathbf{k}}$ by magnitude, color-coded by Fermi surface sheet as in (b).

We now come back to the overestimation of the measured critical temperature in our calculations. At present it is unclear whether this overestimation relates to an inadequate description of anharmonic effects, to the use of the Migdal approximation, to the approximate treatment of retardation effects, or to the assumption of constant DOS near the Fermi level in the Eliashberg theory [24]. In order to test the sensitivity of our results to some of these effects, we repeated our calculations (i) by varying the frequency of the anharmonic modes, and (ii) by varying the DOS via a rigid shift of the Fermi level. Supplemental Fig. S7 [35] shows that our calculated $T_{\rm c}$ is insensitive to the frequency of the anharmonic modes over a wide range, therefore we can exclude this scenario. In contrast, Supplemental Fig. S8 [35] shows that the critical temperature is in better agreement with experiments if we raise the Fermi level by only 200 meV. Motivated by this observation we performed quasiparticle GW calculations of the band structure of $NbS_2$ [26, 27, 52, 55–57]. Supplemental Fig. S9 [35] shows that GW quasiparticle corrections reduce the DOS at the Fermi energy by 18% with respect to our DFT calculations. By repeating our Eliashberg calculations with the corrected $N_{\rm F}$ and $\mu^*$ we find that the critical temperature also decreases by 18%. For completeness the superconducting gap calculated after including quasiparticle corrections is shown in Supplemental Fig. S10 [35]. Our most accurate value, $T_{\rm c} = 15.3$ K, is in better agreement with experiment.

In conclusion, we used the anisotropic *ab initio* Migdal-Eliashberg theory including Coulomb interactions to elucidate the nature of the superconducting pairing in $2H$-$NbS_2$. Our calculations indicate that a large contribution to the pairing comes from EPI hot spots on the triangular Fermi surface arcs, which signal a latent CDW instability. We successfully explained tunneling measurements in terms of two superconducting gaps, a large one associated with in-plane Nb orbitals, and a small one related to the out-of-plane orbitals of Nb and S. More generally, our work highlights the importance of determining accurate low-energy band structures and Fermi surfaces beyond DFT in order to achieve a fully *ab initio* description of the pairing mechanism in TMDs.

We gratefully acknowledge fruitful discussions with M. A. Perez Osorio, M. Zacharias, N. Zibouche, C. Verdi and L. Boeri. This work was supported by the Austrian Science Fund (FWF) Project No. J 3806-N36, the Leverhulme Trust (Grant RL-2012-001), the UK Engineering and Physical Sciences Research Council (grants No. EP/J009857/1 and EP/M020517/1), the European Union's Horizon 2020 programme under grant No. 696656 GrapheneCore1, the University of Oxford Advanced Research Computing (ARC) facility (http://dx.doi.org/810.5281/zenodo.22558) and the ARCHER UK National Supercomputing Service under the 'AMSEC' Leadership project.

---

* present address: King's College London, Physics Department, Strand, London WC2R 2LS, United Kingdom
† feliciano.giustino@materials.ox.ac.uk

# Supplemental Material

# Origin of superconductivity and latent charge density wave in $NbS_2$

Christoph Heil,[1] Samuel Poncé,[1] Henry Lambert,[1,*] Martin Schlipf,[1] Elena R. Margine,[2] and Feliciano Giustino[1,†]

[1] *Department of Materials, University of Oxford, Parks Road, Oxford OX1 3PH, United Kingdom*

[2] *Department of Physics, Applied Physics and Astronomy, Binghamton University-SUNY, Binghamton, New York 13902, USA*

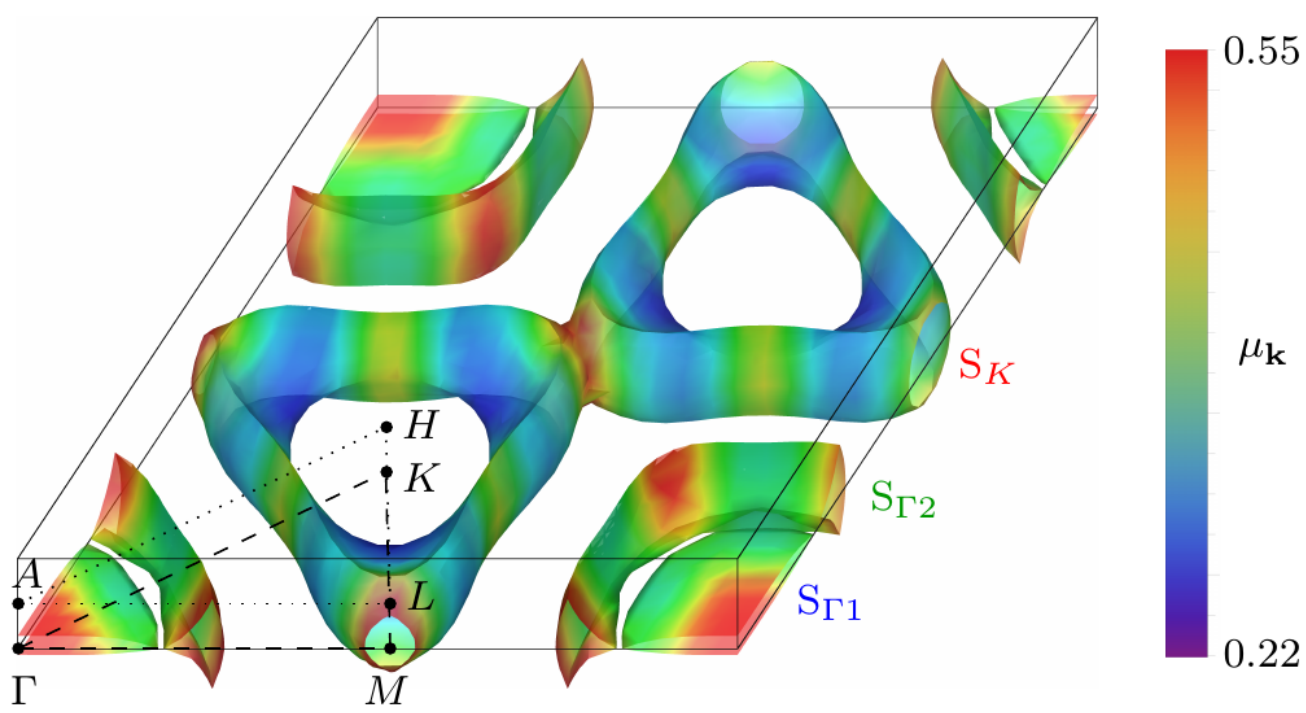

Figure S1. (Color online) Distribution of the pair-breaking screened Coulomb interaction $\mu_k$ on the Fermi surface, with blue indicating the lowest value and red the highest. $\mu_{\mathbf{k}}$ is defined as the Fermi-surface average of the screened Coulomb interaction W: $\mu_{\mathbf{k}} = N_{\mathrm{F}} \langle \langle \mathbf{k}, -\mathbf{k} | W | \mathbf{k}', -\mathbf{k}' \rangle \rangle_{\mathrm{FS}}$.

## ANHARMONIC PHONON CALCULATIONS

Here we describe our approach to determine the vibrational properties of $2H$-$NbS_2$ including anharmonic effects. First, we perform density-functional perturbation theory (DFPT) calculations in the harmonic approximation to obtain the vibrational eigenmodes $\mathbf{e}^{\nu}_{\mathbf{q},\kappa}$ and eigenfrequencies $\omega_{\mathbf{q},\nu}$ of atom $\kappa$ and mode $\nu$ at wavevector $\mathbf{q}$ on a uniform $N_1^p \times N_2^p \times N_3^p$ grid centered at Γ. The corresponding dynamical matrix in the Bloch representation is given by:

$$D^{\mathrm{ph}}_{\mathbf{q},\mu\nu} = \langle \mathbf{q}\mu | \hat{D}^{\mathrm{ph}} | \mathbf{q}\nu \rangle = \delta_{\mu\nu} \omega^2_{\mathbf{q},\nu}. \tag{1}$$

In the phonon Wannier representation the same matrix can be written as [1]:

$$D^{\mathrm{ph}}_{\mathbf{R}_p,\mathbf{R}'_p} = \langle \mathbf{R}_p | \hat{D}^{\mathrm{ph}} | \mathbf{R}'_p \rangle = \sum_{\mathbf{q}} e^{-i\mathbf{q}\cdot(\mathbf{R}'_p - \mathbf{R}_p)} E_{\mathbf{q}} D^{\mathrm{ph}}_{\mathbf{q}} E^{\dagger}_{\mathbf{q}}. \tag{2}$$

Here $\mathbf{R}_p$ denote the lattice vectors and we have rewritten the vibrational eigenvectors as square matrices $(E_{\mathbf{q}})_{\mu\nu}$. The phonon eigenmodes and eigenfrequencies at an arbitrary wavevector $\mathbf{q}'$ can then be obtained by inverting Eq. (2), i.e.:

$$D^{\mathrm{ph}}_{\mathbf{q}'} = E^{\dagger}_{\mathbf{q}'} \left( \frac{1}{N_p} \sum_{\mathbf{R}_p} e^{-i\mathbf{q}'\cdot\mathbf{R}_p} D^{\mathrm{ph}}_{\mathbf{0}_p,\mathbf{R}_p} \right) E_{\mathbf{q}'} \ . \tag{3}$$

In order to calculate the eigenfrequencies and eigenmodes for a wavevector $\mathbf{q}'$, one first has to Fourier interpolate the dynamical matrix in Wannier representation [the term in brackets of Eq. (3)] and then diagonalize the resulting matrix. For further details we refer the reader to Ref. 1.

To calculate phonon dispersion relations in the presence of anharmonicity, we retain the (harmonic) DFPT eigenmodes $(E_{\mathbf{q}})_{\mu\nu}$, and we correct the eigenfrequencies $\omega_{\mathbf{q},\nu}$ in Eq. (1) using calculations for anharmonic potential energy surfaces. The interatomic force constants in Eq. (2) are then computed using the modified frequencies, and the Fourier interpolation in Eq. (3) is finally used to reconstruct the band structures.

The anharmonic corrections to the vibrational frequencies are calculated as follows. We first compute the vibrational eigenmodes for a supercell constructed so as to fold the desired wavevector $\mathbf{q}$ onto the $\Gamma$ point. Then we evaluate the adiabatic potential energy surface (APES) by calculating the total energy of the system as a function of the atomic displacements according to the selected phonon eigenvector, as in the frozen-phonon approach [2]. The atomic displacements are given by:

$$\Delta\tau_{\kappa,\alpha}(x) = \sqrt{\frac{M_0}{M_\kappa}} e_{\kappa,\alpha} x, \tag{4}$$

where $M_\kappa$ is the nuclear mass of atom $\kappa$ and $M_0$ the proton mass. We calculate the eigenvalues of the APES generated by this displacement; this is achieved by solving a one-dimensional Schrödinger equation, following the same procedure outlined in Ref. 3, Sec. 8.2.

As an example, the anharmonic APES for the lowest-energy phonon of $NbS_2$ at $M$ is shown as a solid black line in Fig. S2(a). For comparison we also show the harmonic potential as a dashed red line. The square moduli of the calculated wavefunctions, vertically shifted by their eigenvalue for clarity, are shown as solid colored lines in Fig. S2(a). We performed similar calculations for the wavevectors at $2/3\Gamma M$ and $L$.

To check the sensitivity of $T_c$ to the frequencies of the anharmonic phonons, we have repeated the Migdal-Eliashberg calculations by artificially shifting the frequencies of the anharmonic modes, as shown in Fig. S7. We find that $T_c$ is relatively insensitive to small changes in these frequencies. For example, changing the frequencies of both anharmonic modes by $-4.5$ meV, so as to set the frequency of the first phonon mode at $M$ almost to zero, reduces $T_c$ by only 12%. We also checked that the results are insensitive to the ordering of the lowest two vibrational modes.

The approach just described includes by construction phonon-phonon interactions within the *same* harmonic eigenmode, but it neglects phonon-phonon interactions between *different* eigenmodes. In order to check the magnitude of the inter-mode couplings it is necessary to consider a multi-dimensional APES, whereby the atomic displacements are expanded in terms of multiple normal coordinates:

$$\Delta\tau_{\kappa,\alpha}(x_{\nu_1}, ..., x_{\nu_N}) = \sqrt{\frac{M_0}{M_\kappa}} \left( e^{\nu_1}_{\kappa,\alpha} x_{\nu_1} + ... + e^{\nu_N}_{\kappa,\alpha} x_{\nu_N} \right) \ . \tag{5}$$

Since this operation is computationally prohibitive, we tested the coupling between the two anharmonic phonons at the $M$ point by calculating two-dimensional potential energy surfaces. The square moduli of the first three eigenstates of the 2D-APES are shown in Fig. S2(c)-(e). The calculated phonon frequencies, 4.4 meV and 8.6 meV, are in good agreement with the values that we obtained by using 1D-APES calculations, namely 4.7 meV and 8.1 meV respectively. This favorable comparison indicates that for the purposes of the present work it is legitimate to consider 1D-APES calculations of anharmonic corrections.

When compared to perturbative calculations of anharmonicity, the approach that we employ here carries the advantage that anharmonicity (within a single eigenmode) is included to all orders. Furthermore, when used in conjunction with standard Fourier interpolation techniques of the interatomic force constants, this method can provide complete phonon dispersion relations. The inclusion of inter-mode anharmonic coupling is also possible by considering multi-dimensional APES, although these calculations can become very time-consuming.

The disadvantages of the method used here are: (i) the procedure is limited to $\mathbf{q}$-vectors which are commensurate with small supercells (most typically high-symmetry points in the Brillouin zone); (ii) inter-mode anharmonic phonon-phonon interactions can only be considered for a small subset of phonon modes.

In contrast, self-consistent approaches to anharmonic phonons [4–8] automatically incorporate both intra-mode and inter-mode anharmonic couplings. In addition, these approaches allows for the incorporation of temperature in the calculated phonon spectra [7]. Therefore self-consistent methods are more general than the present approach, but come at an increased computational cost.

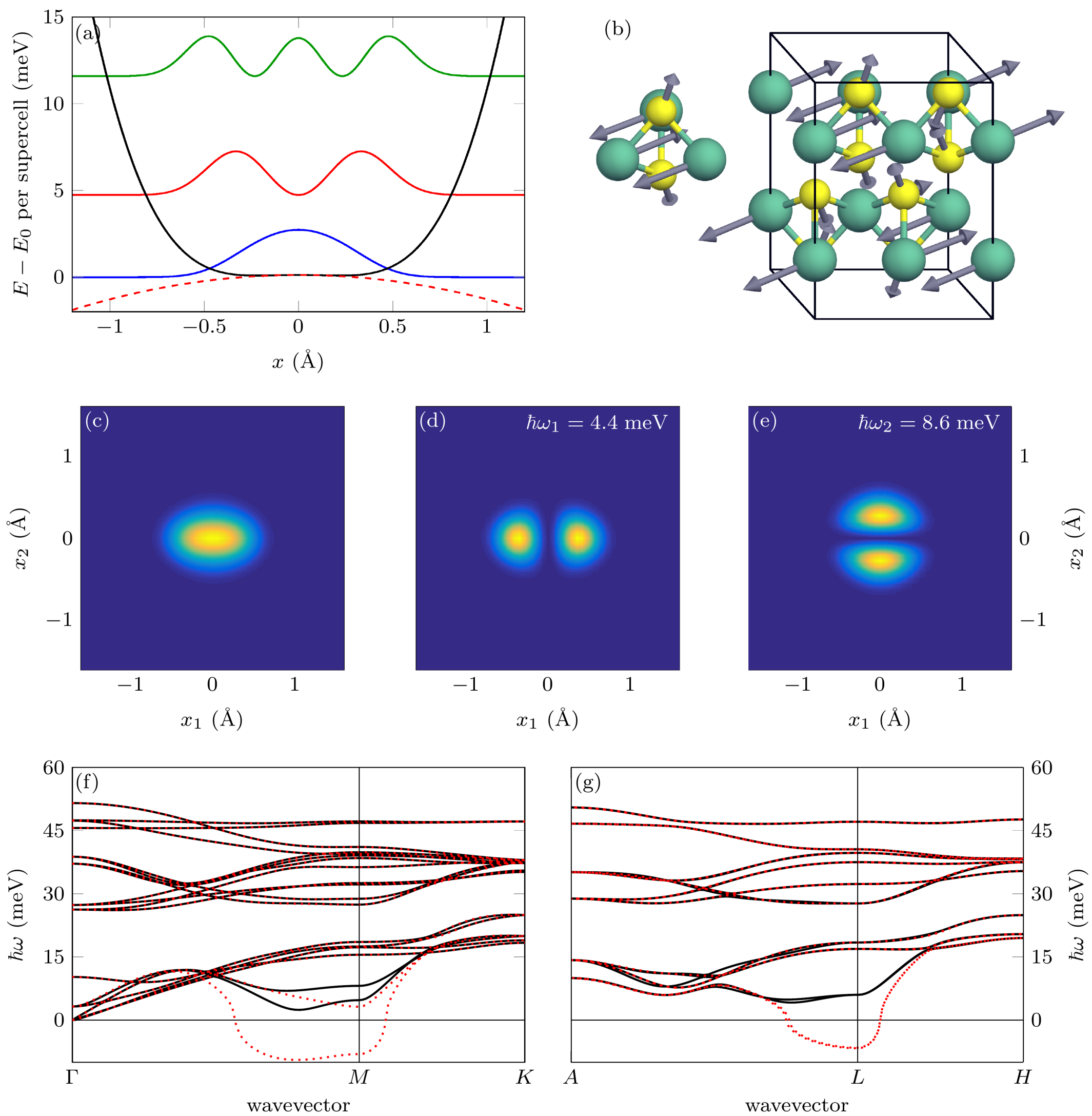

Figure S2. (Color online) (a) Adiabatic potential energy surface (APES) of $2H$-$NbS_2$, calculated along the normal mode coordinate $x$ corresponding to the lowest-energy mode at $M$, fitted up to $8^{\text{th}}$ order (solid black line). The dashed red line shows the harmonic part of the APES. The square moduli of the wavefunctions corresponding to the APES in panel (a) are shown in order of increasing energy as solid colored lines. These curves have been vertically shifted by their energy eigenvalue. The calculation corresponding to panel (a) was carried out for a $2 \times 1 \times 1$ supercell so as to fold $M$ into Γ, with a uniform $18 \times 36 \times 12$ **k**-grid and an electronic smearing parameter of 0.005 Ry. We performed similar calculations for the phonons at $2/3\,\Gamma M$ and at $L$, obtaining similar results. (b) Atomic displacement pattern corresponding to the lowest-energy phonon at $M$, showing the vibration of the $Nb_3S_2$ trigonal bipyramidal units. (c)-(e) Square moduli of the first three wavefunctions calculated for the 2-dimensional APES (again fitted up to $8^{\text{th}}$ order in both coordinates), taking into account the coupling between the two anharmonic phonon modes at $M$. The calculated eigenfrequencies are in good agreement with the uncoupled, 1-dimensional APES calculations (4.4 and 8.6 meV vs. 4.7 and 8.1 meV, respectively). (f)-(g) Comparison between the phonon dispersion relations of $2H$-$NbS_2$ calculated within the harmonic approximation (dotted red lines), and using fully anharmonic calculations (solid black lines) as described in the previous section.

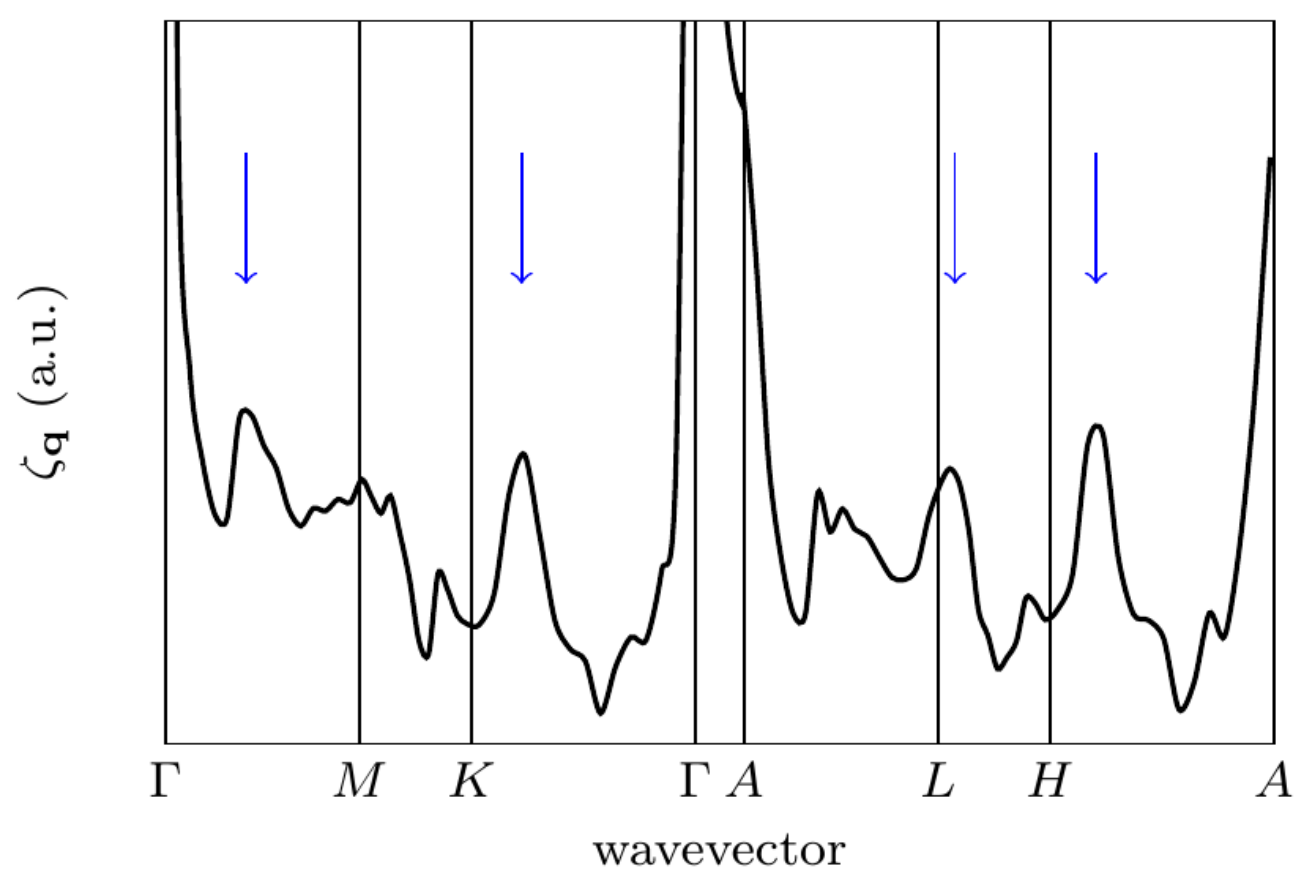

Figure S3. (Color online) Calculated Fermi surface nesting function $\zeta_{\mathbf{q}}$ of $2H$-$NbS_2$ along a high-symmetry path in the Brillouin zone, where the arrows point to local maxima. For a definition of the nesting function see Ref. 9 and 10. Peaks in this function can be used to identify nesting vectors, which connect parallel sheets of the Fermi surface (apart from $\mathbf{q} = 0$, where the peak is an artifact of the definition). In this case, we do not find any clear indication of strong nesting inducing large EPI. The nesting function was calculated using a $200{\times}200{\times}100$ Brillouin grid and a Gaussian smearing of 25 meV.

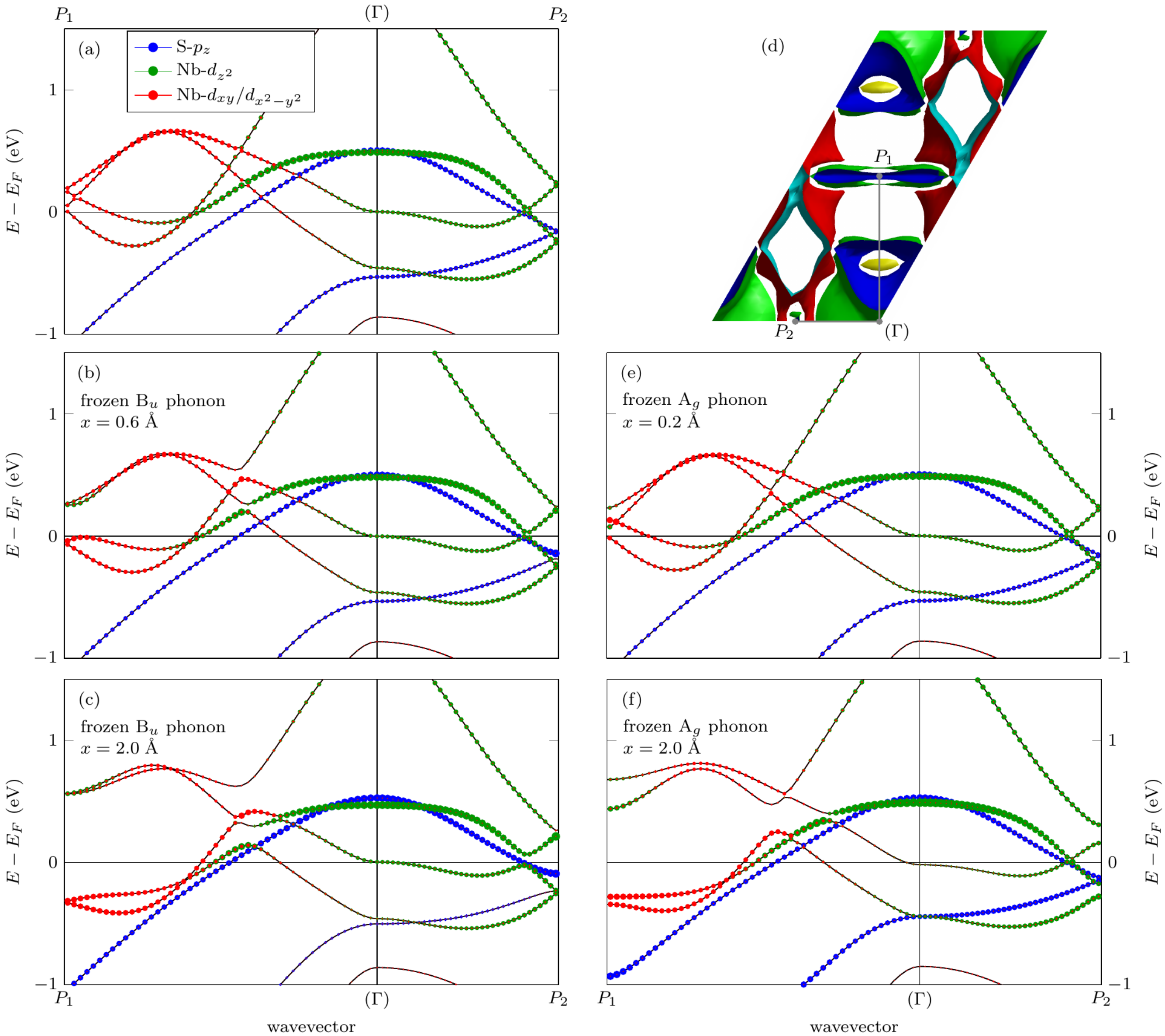

Figure S4. (Color online) (a)-(c) Calculated band structure of $2H$-$NbS_2$ in a $2 \times 1 \times 1$ supercell. In (a) we show the band structure in the ground state, in (b) the band structure calculated after displacing the atoms according to the $B_u$ phonon mode at $M$, so as to place the structure in one of the minima of the APES; and in (c) the band structure for a normal mode coordinate displacement of $x = 2$ Å, which corresponds to an absolute displacement of the Nb atoms of 0.1 Å (see previous section on the anharmonic phonon calculations for a definition of the normal mode coordinate). We see that the atomic displacement induces an avoided crossing in the band structure around the $P_1$ point with coordinates $(0.51, 0.88, 0.00)\ 2\pi/a$. The splitting of electronic degeneracies and the removal of electronic states around the Fermi energy (compare with Fig. S5) by a linear electron-phonon coupling is analogous to the dynamical Jahn-Teller effect in molecules [11], and is associated with a phonon softening and a latent lattice instability [12, 13]. (d) Calculated Fermi surfaces corresponding to the band structure in (a). The vibration of the $B_u$ phonon mode at $M$ modulates the shape of the Fermi surface, as it is seen in Fig. 3 of the main text. (e)-(f) Similar to (b)-(c), but the atoms have been displaced according to the $A_g$ phonon mode.

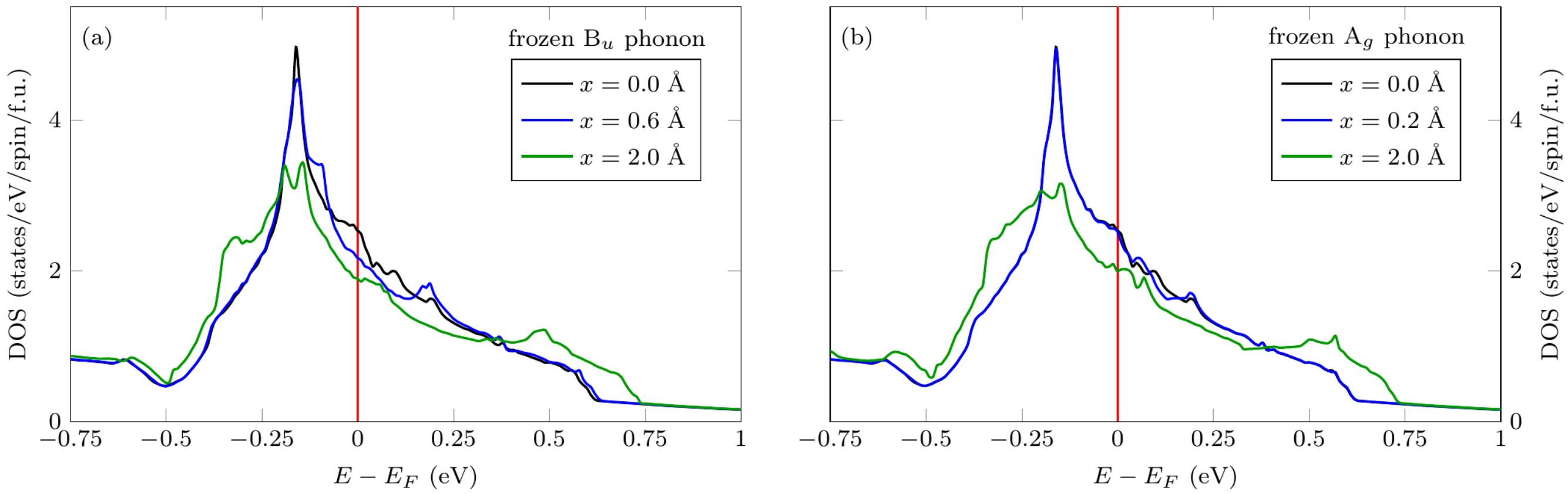

Figure S5. (Color online) (a)-(b) Calculated DOS of $2H$-$NbS_2$ in a $2 \times 1 \times 1$ supercell. In (a) we show the DOS in the ground state (black), and after displacing the atoms according to the $B_u$ phonon mode at $M$ for displacements of $x = 0.6$ Å [blue; see also Eq. (4)], which corresponds to placing the structure in the minimum of the anharmonic double-well potential, and $x = 2.0$ Å (green), which corresponds to a displacement of the Nb atoms by 0.1 Å. The DOS of the ground state structure exhibits a shoulder close to the Fermi energy, which is removed when the atoms are displaced along the $B_u$ mode. (b) Similar to (a), but the atoms have been displaced according to the $A_g$ phonon mode, with displacements of $x = 0.2$ Å (blue) and $x = 2.0$ Å (green). We observe a similar behavior as in (a), but not as pronounced. These figures show that the splitting of electronic degeneracies and the removal of electronic states from the Fermi level (compare with Fig. S4) induce significant changes in the electronic structure of $2H$-$NbS_2$ and play a role in the phonon softening in this compound [12, 13].

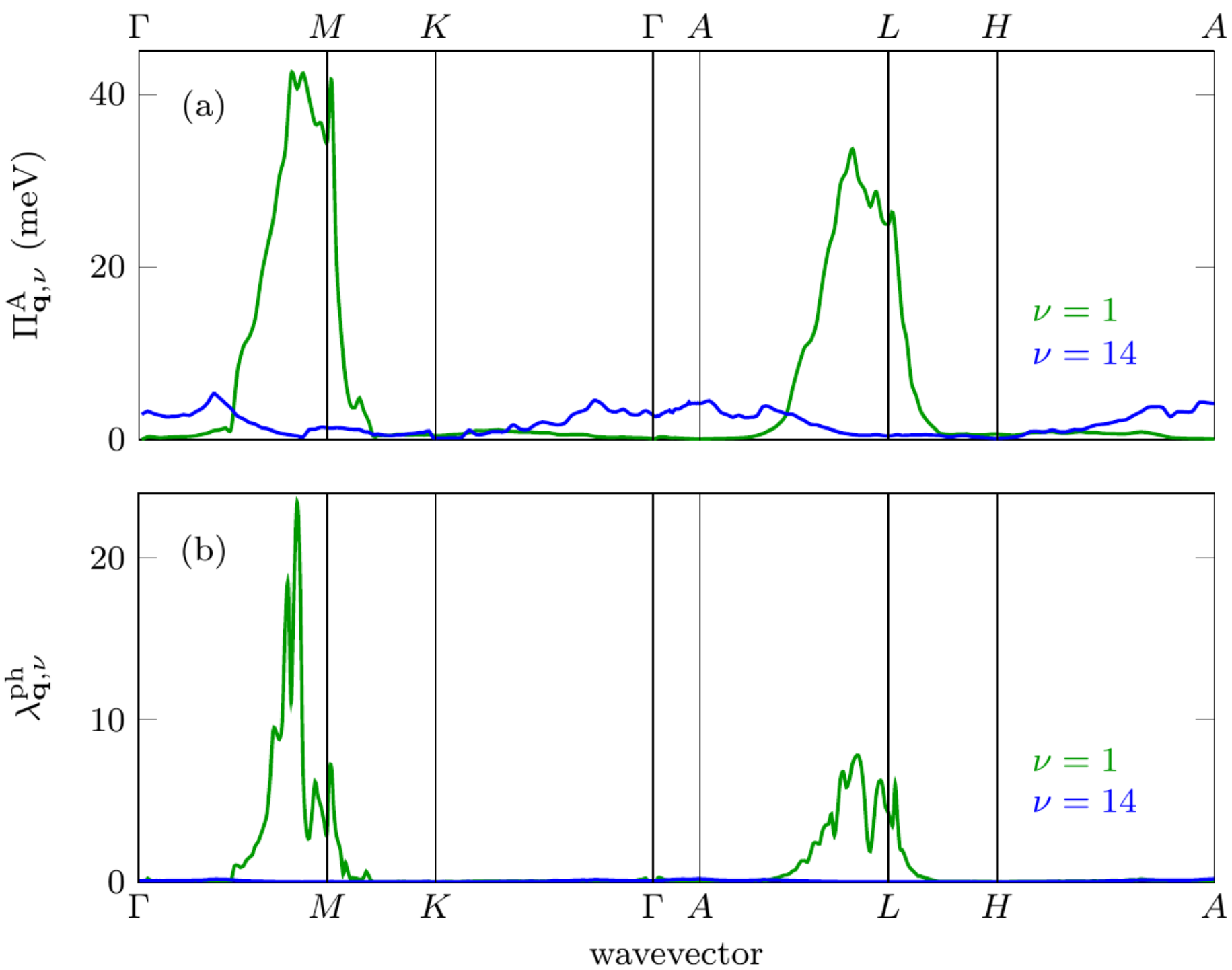

Figure S6. (Color online) (a) The Fermi surface contribution to the adiabatic phonon self-energy [14]
$\Pi^{\mathrm{A}}_{\mathbf{q},\nu} = \sum_{m,n} \int \frac{d\mathbf{k}}{\Omega_{\mathrm{BZ}}} |g_{mn,\nu}(\mathbf{k},\mathbf{q})|^2 \left[\frac{f_{n,\mathbf{k}} - f_{m,\mathbf{k}+\mathbf{q}}}{\epsilon_{m,\mathbf{k}+\mathbf{q}} - \epsilon_{n,\mathbf{k}}}\right] \Theta(|\epsilon_{m,\mathbf{k}+\mathbf{q}} - \epsilon_F| < \epsilon_c)\Theta(|\epsilon_{n,\mathbf{k}} - \epsilon_F| < \epsilon_c)$,
with electron band indices $m, n$, phonon mode index $\nu$, reciprocal wavevectors $\mathbf{k}, \mathbf{q}$, electron-phonon matrix elements $g_{mn,\nu}$, Fermi distribution function $f$, Brillouin zone volume $\Omega_{\mathrm{BZ}}$ and Fermi energy $\epsilon_F$. The step-function $\Theta$ with an energy cut-off $\epsilon_c = 100\,\mathrm{meV}$ ensures that only electronic states close to $\epsilon_F$ are contributing to the self-energy, the qualitative results are insensitive to the choice of $\epsilon_c$. We observe that $\Pi^{\mathrm{A}}_{\mathbf{q},\nu}$ for the first phonon mode (green) has large values in the vicinity of $2/3\Gamma M$ ($2/3AL$) and is close to zero everywhere else, while $\Pi^{\mathrm{A}}_{\mathbf{q},\nu}$ for the 14$^{\mathrm{th}}$ phonon mode (blue) has a much lower and uniform response along the high-symmetry path. (b) EPI strength $\lambda^{\mathrm{ph}}_{\mathbf{q},\nu}$ for the same two phonon modes as in (a). The fact that $\Pi^{\mathrm{A}}_{\mathbf{q},\nu}$ is largest for those wavevectors $\mathbf{q}$ and phonon modes $\nu$ that also exhibit the largest EPI $\lambda^{\mathrm{ph}}_{\mathbf{q},\nu}$ provides further evidence that the observed phonon softening is due to the large EPI associated with the anharmonic phonon modes in the vicinity of $2/3\Gamma M$ ($2/3AL$).

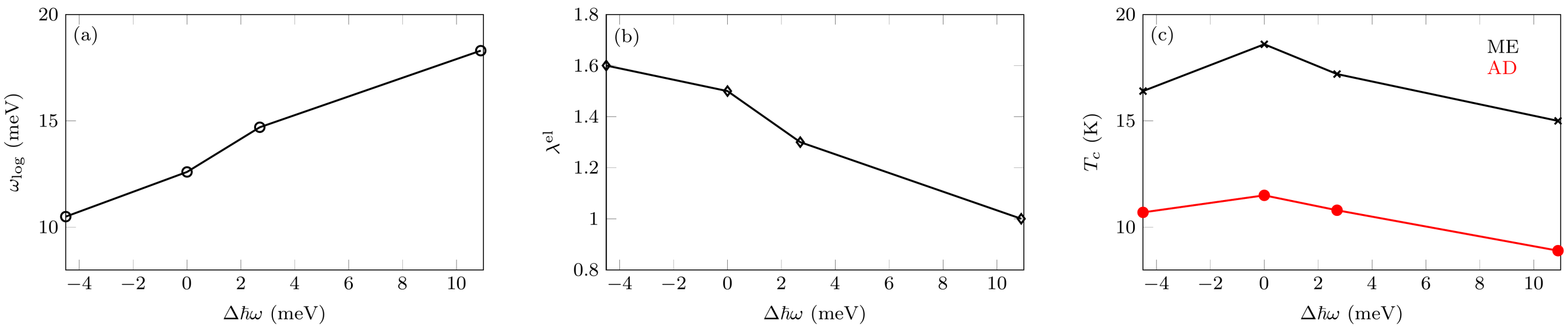

Figure S7. (Color online) Sensitivity of the calculated superconducting properties of $2H$-$NbS_2$ to the energy of the anharmonic modes. In panel (a) we show the logarithmic average phonon energy $\omega_{\rm log}$ and in panel (b) the total EPI strength $\lambda^{\rm el}$ as a function of the energy shift $\Delta\hbar\omega$. In panel (c) we show the superconducting critical temperature calculated using the anisotropic Migdal-Eliashberg theory ('ME', black) and the Allen-Dynes formula ('AD', red) as a function of the energy shift $\Delta\hbar\omega$. In order to perform these calculations we rigidly shifted the energies of all the soft modes by the amount indicated on the horizontal axis. The critical temperature is found to be relatively insensitive to the phonon frequencies, owing to the compensating variations in $\omega_{\rm log}$ and $\lambda^{\rm el}$.

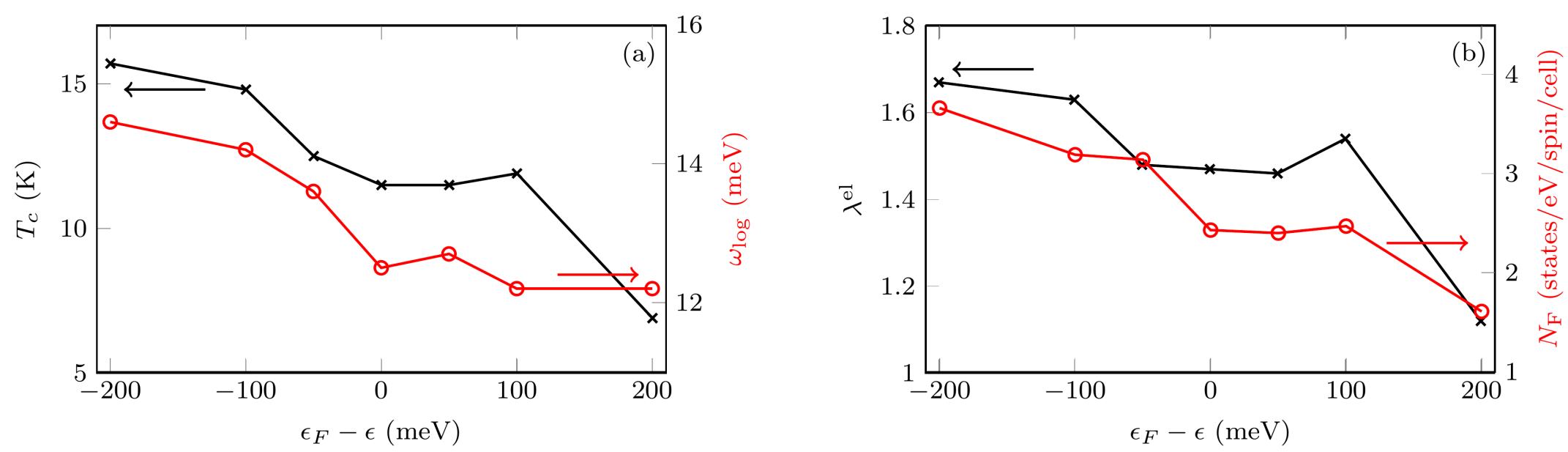

Figure S8. (Color online) Sensitivity of the calculated superconducting properties of $2H$-$NbS_2$ to the Fermi energy. (a) Variation of $T_{\rm c}$ calculated using the Allen-Dynes formula (black crosses) and of the logarithmic average phonon energy $\omega_{\rm log}$ (red circles) as a function of a rigid shift of the Fermi level. (b) Variation of the EPI strength $\lambda^{\rm el}$ (black crosses) and the DOS at the Fermi level ('$N_{\rm F}$', red circles) as a function of a rigid shift of the Fermi level. The large variation of the DOS with the Fermi energy has a significant impact on the calculated $T_{\rm c}$, mostly due to its effect on the EPI strength.

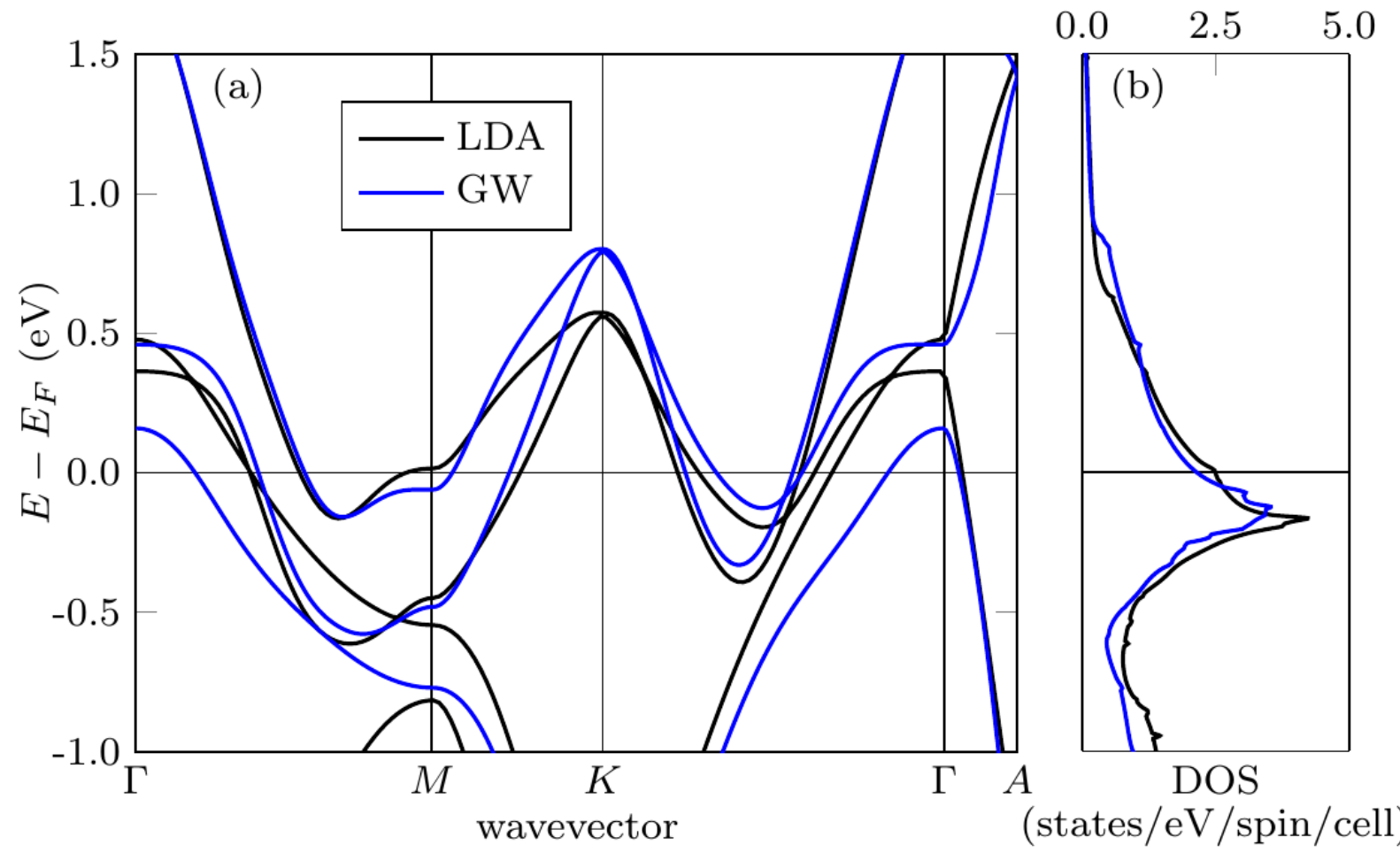

Figure S9. (Color online) Comparison between calculations of the band structure (a) and density of states (b) of $NbS_2$ within the local density approximation to DFT (black) and within the $G_0W_0$ approximation as obtained using the SternheimerGW code (blue) [15, 16]. The $G_0W_0$ calculations were performed using the Godby-Needs plasmon pole approximation [17], using an imaginary pole energy of 4 eV. The frequency integration was performed along the imaginary axis, and the self-energy on the real axis was obtained using analytic continuation based on Padé approximants of order 11 [15, 16]. For the screened Coulomb interaction we used a 12×12×4 Brillouin zone grid, and we determined the quasiparticle corrections to the Kohn-Sham eigenvalues on a 6×6×2 Brillouin zone grid. These data were subsequently interpolated onto a fine 60×60×20 Brillouin zone grid by means of maximally localized Wannier functions [18, 19]. We employed an energy cutoff of 10 Ry for the dielectric matrix, and an exchange self-energy cutoff of 25 Ry. We note that the SternheimerGW method avoids the calculation of unoccupied electronic states, therefore no convergence over such states is required. The quasiparticle corrections decrease the DOS at the Fermi level by 18%.

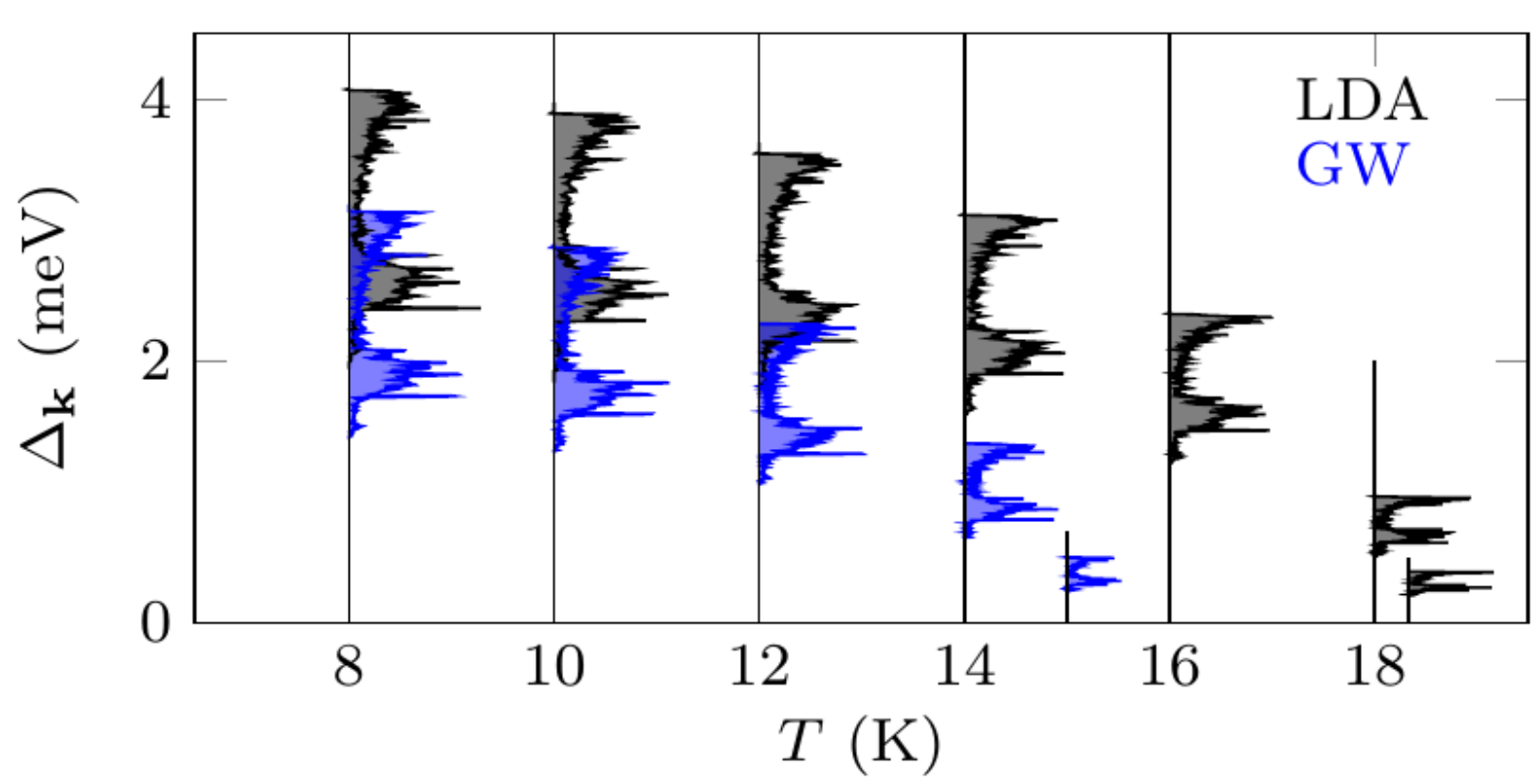

Figure S10. (Color online) Calculated energy distribution of the superconducting gap $\Delta_{\mathbf{k}}$ as a function of temperature. The black curves (LDA) correspond to the result shown in Fig. 4(a) of the main text, the blue curves (GW) take into account the change in the DOS at the Fermi level due to GW quasiparticle corrections, as described in Fig. S8. The quasiparticle corrections lower $T_c$ by 18%, from 18.6 K (LDA) to 15.3 K (GW). Similarly the superconducting gap $\Delta_{\mathbf{k}}(T \to 0)$ decreases from $\sim$ 4.2 meV to $\sim$ 3.3 meV.

---

* present address: King's College London, Physics Department, Strand, London WC2R 2LS, United Kingdom
† feliciano.giustino@materials.ox.ax.uk